\documentclass[aps,prl,twocolumn,superscriptaddress]{revtex4}
\usepackage{graphicx}
\usepackage{epstopdf}
\usepackage{color}

\begin{document}

\title{Two-dimensional Massless Dirac Fermions in Antiferromagnetic \emph{A}Fe$_{2}$As$_{2}$ (\emph{A} = Ba, Sr)}

\author{Zhi-Guo Chen} \email{zgchen@iphy.ac.cn}
\affiliation{Beijing National Laboratory for Condensed Matter Physics, Institute of Physics,
Chinese Academy of Sciences, Beijing 100190, China}

\author{Luyang Wang}
\affiliation{Institute for Advanced Study, Tsinghua University, Beijing 100084, China}
\affiliation{Sate Key Laboratory of Optoelectronic Materials and Technologies, School of Physics, Sun Yat-Sen University, Guangzhou 510275, China}

\author{Yu Song}
\affiliation{Department of Physics and Astronomy, Rice University, Texas 77005, USA}

\author{Xingye Lu}
\affiliation{Beijing National Laboratory for Condensed Matter Physics, Institute of Physics, Chinese Academy of Sciences, Beijing 100190, China}

\author{Huiqian Luo}
\affiliation{Beijing National Laboratory for Condensed Matter Physics, Institute of Physics,
Chinese Academy of Sciences, Beijing 100190, China}

\author{Chenglin Zhang}
 \affiliation{Department of Physics and Astronomy, Rice University, Texas 77005, USA}

\author{Pengcheng Dai}
\affiliation{Department of Physics and Astronomy, Rice University, Texas 77005, USA}

\author{Zhiping Yin} \email{yinzhiping@bnu.edu.cn}
\affiliation{Center of Advanced Quantum Studies and Department of Physics, Beijing Normal University, Beijing 100875, China}

\author{Kristjan Haule}
\affiliation{Department of Physics and Astronomy, Rutgers University, New Jersey 08854, USA}
\affiliation{DMFT-MatDeLab Center, Upton, New York 11973, USA}

\author{Gabriel Kotliar}
\affiliation{Department of Physics and Astronomy, Rutgers University, New Jersey 08854, USA}
\affiliation{DMFT-MatDeLab Center, Upton, New York 11973, USA}

\begin{abstract}
	We report infrared studies of \emph{A}Fe$_{2}$As$_{2}$ (\emph{A} = Ba, Sr), two representative parent compounds of iron-arsenide superconductors, at magnetic fields (\emph{B}) up to 17.5 T. Optical transitions between Landau levels (LLs) were observed in the antiferromagnetic states of these two parent compounds. Our observation of a $\sqrt{B}$ dependence of the LL transition energies, the zero-energy intercepts at \emph{B} = 0 T under the linear extrapolations of the transition energies and the energy ratio ($\sim$ 2.4) between the observed LL transitions, combined with the linear band dispersions in two-dimensional (2D) momentum space obtained by theoretical calculations, demonstrates the existence of massless Dirac fermions in antiferromagnetic BaFe$_{2}$As$_{2}$. More importantly, the observed dominance of the zeroth-LL-related absorption features and the calculated bands with extremely weak dispersions along the momentum direction $k_{z}$ indicate that massless Dirac fermions in BaFe$_{2}$As$_{2}$ are \textit{2D}. Furthermore, we find that the total substitution of the barium atoms in BaFe$_{2}$As$_{2}$ by strontium atoms not only maintains 2D massless Dirac fermions in this system, but also enhances their Fermi velocity, which supports that the Dirac points in iron-arsenide parent compounds are topologically protected. 

\end{abstract}


\maketitle

In condensed matters, massless Dirac fermions (MDF), which represent a type of quasi-particles with linear energy-momentum dispersions, provide a cornerstone for various quantum phenomena, such as novel quantum Hall effect\cite{Geim1,Geim2,PKim}, Klein tunnelling\cite{Beenakker} and giant linear magneto-resistance\cite{NPOng}. Due to their crucial role in diverse quantum phenomena, searching for MDF in new systems is one of the most active research areas in condensed matter science. Generally, massless Dirac fermions with their valence and conduction band touching at a degenerate point (i.e. Dirac point) are protected by symmetries in solids\cite{Kane,SCZhang,YLChen,Hasan,Borisenko}. However, the formation of magnetic order is always accompanied with time-reversal symmetry breaking and sometimes followed by crystalline symmetry lowering. Therefore, it is a challenge to achieve MDF in magnetic ground states of solid-state electronic systems \cite{Park,SCZhang1}. Moreover, two-dimensional (2D) MDF, which were realized in 2D materials and on the surfaces of three-dimensional (3D) topological insulators \cite{Geim1,Geim2,PKim,XJZhou,Hsieh,ZXShen,QKXue,Takagi}, have rarely been observed in the bulk of 3D crystals \cite{Park}. The discovery of iron-arsenide superconductors not only brings people a new class of unconventional superconductivity \cite{Hosono}, but also sheds light on seeking 2D MDF in magnetic compounds\cite{YRan,Tohyama}.

The parent compounds of iron-arsenide superconductors (PCIS) exhibit collinear antiferromagnetic (AFM) order at low temperature (\emph{T})\cite{WeiBao,PDai1}. After the AFM phase transition, the paramagnetic Brillouin zone of PCIS would be folded along the AFM wave vector connecting the hole-type Fermi surfaces (FSs) surrounding the $\Gamma$ point and the electron-type FSs at the M point, which leads to the intersection between the electron- and hole-bands and results in the bandgap opening at the band crossing points near Fermi energy (\emph{E}$_{F}$) \cite{nlwang1,nlwang2,CCHomes,Uchida,Basov}.
However, theoretical studies suggest that for PCIS, some band crossing points are \textit{topologically} protected by (i) the collinear AFM order, (ii) inversion symmetry and (iii) a combination of time-reversal and spin-reversal symmetry \cite{YRan,Tohyama}.
Furthermore, angle-resolved photoemission spectroscopy (ARPES) measurements of BaFe$_{2}$As$_{2}$ show linear band dispersions within $k_x$-$k_y$ plane below \emph{E}$_{F}$ \cite{Richard}. Therefore, massless fermions, which are characterized by band-degeneracy points and linear energy-momentum dispersions, are expected to exist in the AFM state of BaFe$_{2}$As$_{2}$ \cite{YRan,Tohyama,Uchida,Harrison,Richard,Tanigaki1,Uji,Karpinski,Maeda,Sugai}. There are three types of massless fermions, including MDF, massless Kane fermions and Weyl fermions \cite{Orlita,Teppe,Orlita1,Hasan1,Weng,Hasan2,HDing,YLChen1}. Therein, massless Kane fermions and Weyl fermions are characterized by a heavy-hole valence band around the band-degeneracy point and pairs of degenerate nodes with opposite chirality, respectively\cite{Orlita,Teppe,Orlita1,Hasan1,Weng}, while in PCIS, the heavy-hole valence band is absent and the degenerate nodes have the same chirality\cite{YRan,Tohyama,Richard,Uji}, which preclude the association of the observed linear band dispersions by ARPES with massless Kane fermions and Weyl fermions. Therefore, massless fermions in PCIS were suggested to be MDF \cite{YRan,Tohyama,Uchida,Harrison,Richard,Tanigaki1,Uji,Karpinski,Maeda,Sugai}. However, whether MDF in PCIS are 2D or 3D remains unclear. Further studies are needed to clarify the dimensionality of MDF in PCIS. Besides, in order to verify whether the degenerate points of PCIS are topologically protected, it is important to experimentally test the robustness of the band crossing points in this system. Experimental investigations of the optical transitions between Landau levels (LLs), combined with band-structure calculations, can provide strong evidence for identifying the nature of MDF in PCIS.

Infrared spectroscopy is a bulk-sensitive experimental technique for investigating low-energy excitations of a material. Here, to study the LL transitions in PCIS, we measured the infrared reflectance spectra \emph{R}(\emph{B}) of \emph{A}Fe$_{2}$As$_{2}$ (\emph{A} = Ba, Sr) single crystals in their AFM states (\emph{T} $\sim$ 4.5 K) with the magnetic field applied along the wave vector of the incident light and the crystalline $\emph{c}$-axis (i.e. Faraday geometry) \cite{Suppl}. Landau-level-transition features are present in the relative reflectance spectra \emph{R}(\emph{B})/\emph{R}($B_0$ = 0 T) of these two PCIS. Our observation of a $\sqrt{B}$ dependence of the LL transition energies, the zero-energy intercepts at \emph{B} = 0 T under the linear extrapolations of the transition energies, the energy ratio ($\sim$ 2.4) between the two measured LL transitions and the dominant absorption features of the zeroth-LL-related transitions, combined with the linear band dispersions in 2D momentum space calculated using the density functional theory and dynamical mean field theory (DFT+DMFT), indicates \textit{2D} MDF in AFM BaFe$_{2}$As$_{2}$. Moreover, to test the robustness of MDF in this system, we completely replaced the barium atoms in BaFe$_{2}$As$_{2}$ with strontium atoms, causing a distortion of the crystal structure \cite{Johnston}. Our investigations show that 2D MDF not only exist in SrFe$_{2}$As$_{2}$, but also have larger effective Fermi velocity after substituting the barium atoms, which provides evidence for the topologically protected Dirac points in PCIS.

\begin{figure}
	\includegraphics[width=8.5cm]{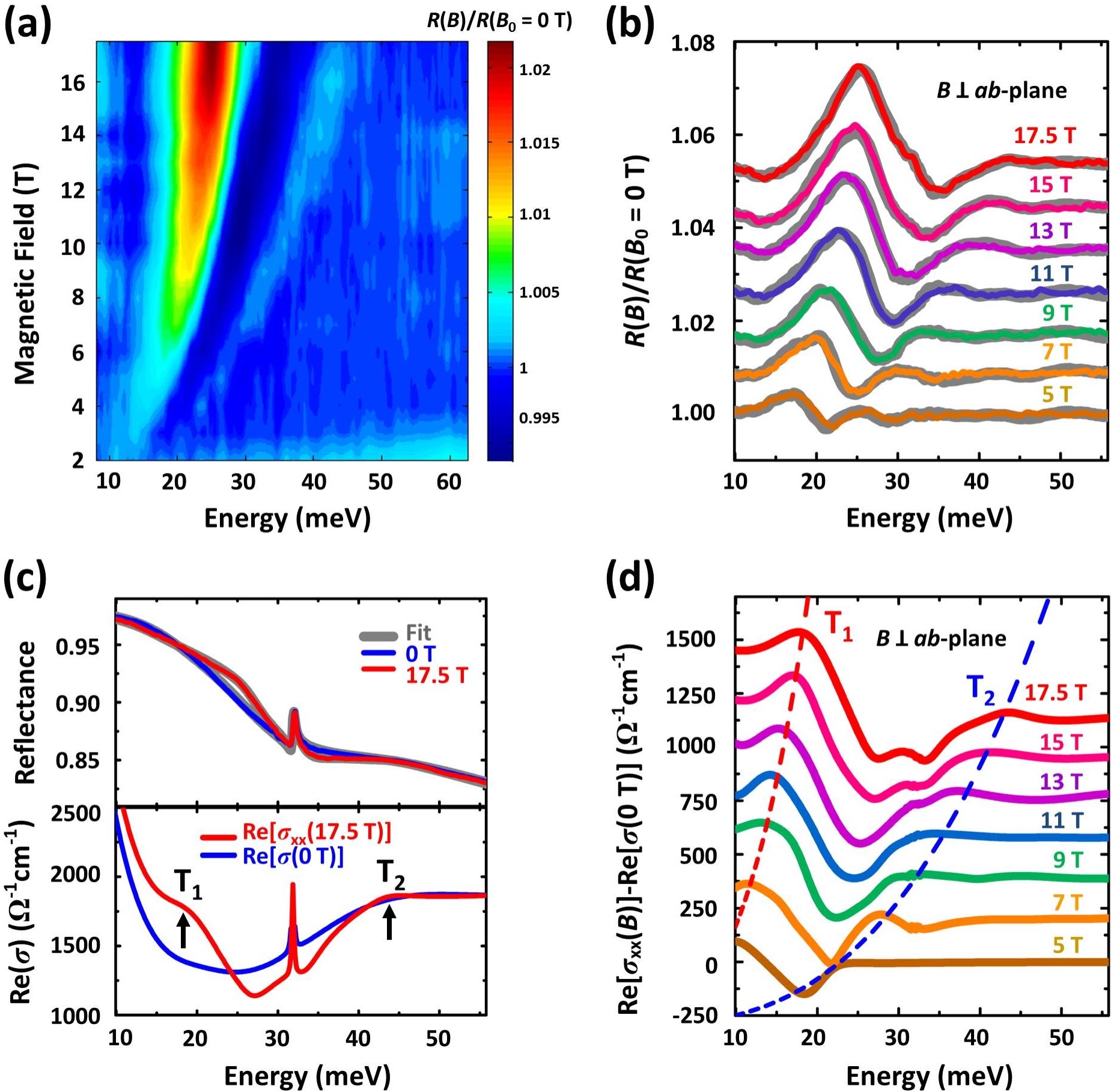}\\
	\centering
	\caption{
		Magneto-optical response of BaFe$_{2}$As$_{2}$. (a) False-color map of the \emph{R}(\emph{B})/\emph{R}($B_0$ = 0 T) spectra. See the false-color map plotted as a function of $\sqrt{B}$ and energy in Supplementary Fig. 4. (b) Several representative \emph{R}(\emph{B})/\emph{R}($B_0$ = 0 T) spectra and their fitting results (grey curves). The \emph{R}(\emph{B})/\emph{R}($B_0$ = 0 T) spectra up to 120 meV are shown in Fig. S5 of the Supplemental Material \cite{Suppl} Fig. 5. (c) Upper panel: the reflectance spectra at \emph{B} = 0 T and \emph{B} = 17.5 T, and their fitting results. Lower panel: the real part of $\sigma$$_{xx}$(\emph{B}, $\omega$) at \emph{B} = 17.5 T and $\sigma$(0 T, $\omega$). The two black arrows in Re[$\sigma$$_{xx}$(\emph{B} = 17.5 T, $\omega$)] indicate two additional peaks, T$_{1}$ and T$_{2}$. (d) Relative real part of the optical conductivity. The blue and red dashed lines are guides to eyes. Except T$_{1}$ and T$_{2}$, another peak emerges around 31 meV in (b) and (d) at \emph{B} $>$ 13 T (see its possible origin in the Supplemental Material \cite{Suppl}, which includes Refs. \cite{Marek1,Marek2,Boeri}). The spectra in (b) and (d) are displaced from one another by 0.09 and 240 ($\Omega$$^{-1}$ cm$^{-1}$) for clarity.
	}
	\label{Fig:linear}
\end{figure}

Figure 1(a) depicts the false-color map of the \emph{R}(\emph{B})/\emph{R}($B_0$ = 0 T) spectra of BaFe$_{2}$As$_{2}$ as a function of magnetic field and energy. As displayed in Fig. 1(b), peak-like features are present in \emph{R}(\emph{B})/\emph{R}($B_0$ = 0 T) and systematically shift to higher energies as the magnetic field increases. A fundamental question is what these peak-like features arise from. Largely due to the strong (next) nearest neighbor exchange coupling\cite{PDai2}, an ultra-high magnetic field (\emph{B} $>$ 500 T) along the crystalline \emph{c}-axis is estimated to destroy the low-temperature collinear AFM order in PCIS\cite{Tokunaga}. The applied magnetic field (\emph{B} $\leq$ 17.5 T) here is therefore too low to dramatically change the AFM order at \emph{T} $\sim$ 4.5 K. In addition, the AFM-order-related spectral-weight-transfer in the reflectance spectra of BaFe$_{2}$As$_{2}$ takes place in the energy range up to about 250 meV\cite{nlwang1,nlwang2,CCHomes,Uchida,Basov}, which is much wider than that observed in Fig. 1 (b) ($\sim$ 55 meV). Based on the applied magnetic field and energy range of the spectral-weight-transfer here, the observed peak-like features are unlikely to originate from the magnetic-field-induced change in the AFM order. Besides, structural twins form in the orthorhombic crystal of BaFe$_{2}$As$_{2}$ below the AFM phase transition temperature\cite{JQLi,Prozorov}. The magnetic field applied along the orthorhombic \emph{a-} or \emph{b-}axis can be used for partially detwinning BaFe$_{2}$As$_{2}$ single crystals\cite{Fisher}. Here, the magnetic field applied perpendicular to the \emph{ab}-plane cannot detwin the single crystals, so the observed peak features in Fig. 1(b) are irrelevant with the detwinning\cite{Dressel}. Given the magnetic-field-dependent energy positions, the peak features should arise from the optical absorption of the LL transitions in BaFe$_{2}$As$_{2}$.

\begin{figure}
	\includegraphics[width=8.6cm]{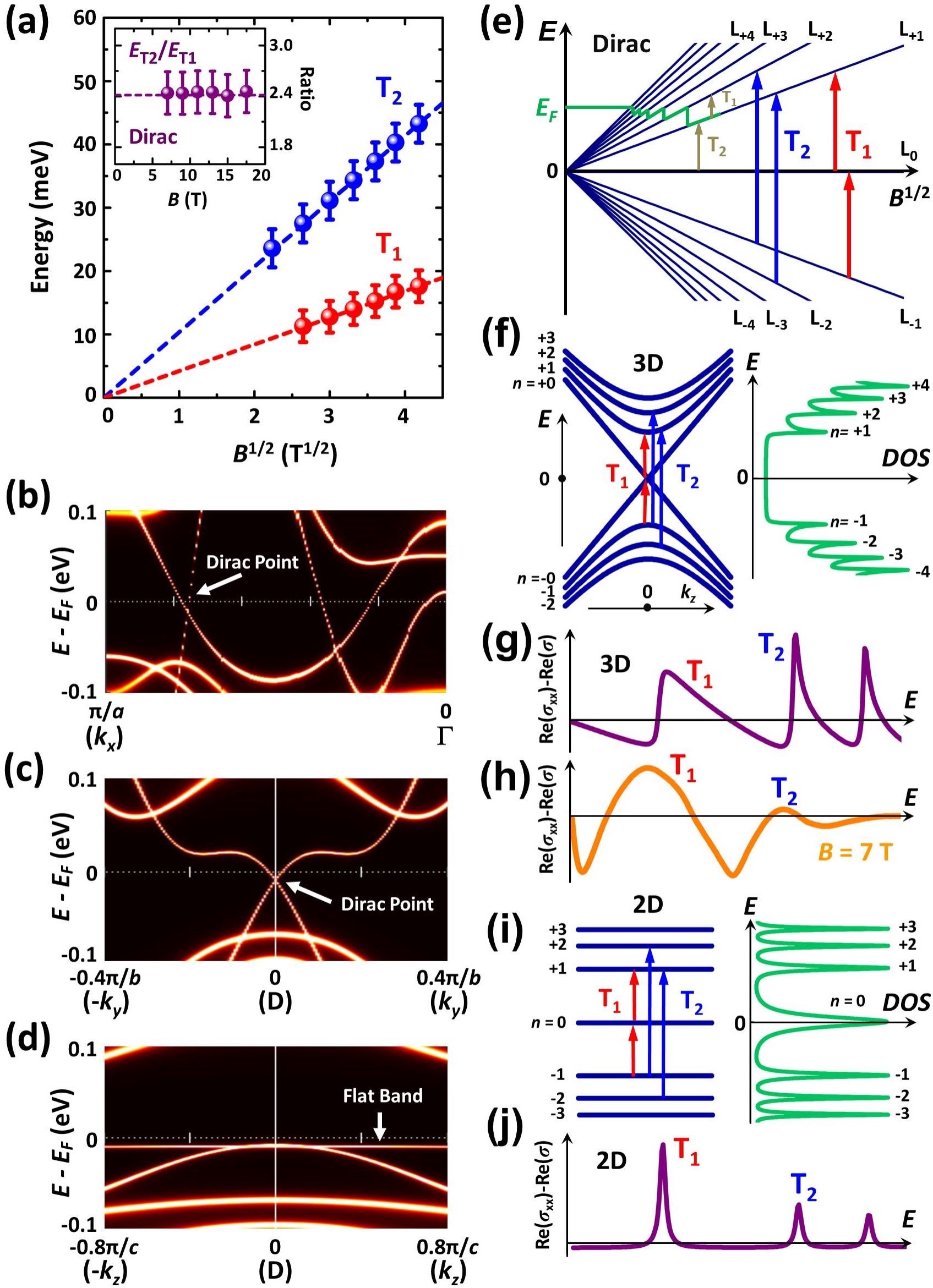}\\
	\centering
	\caption{Landau level transitions and band dispersions of BaFe$_{2}$As$_{2}$ in the AFM state. (a) Observed LL transitions, T$_{1}$ and T$_{2}$, in an \emph{E}-$\sqrt{B}$ plot. The inset shows the energy ratio of T$_{1}$ and T$_{2}$ as a function of \emph{B}. The purple dashed line shows the theoretical energy ratio based on MDF model. (b-d) Calculated band dispersions along $k_x$, $k_y$ and $k_z$ near $E_F$. The flat band in (d) represents the dispersion of MDF along $k_z$. In the AFM Brillouin zone of \textit{A}Fe$_{2}$As$_{2}$ (\textit{A} = Ba, Sr), -$\pi$/\textit{a} $\leq$ $k_x$ $\leq$ $\pi$/\textit{a}, -$\pi$/\textit{b} $\leq$ $k_y $ $\leq$ $\pi$/\textit{b} and -2$\pi$/\textit{c} $\leq$ $k_z$ $\leq$ 2$\pi$/\textit{c}, where \textit{a}, \textit{b} and \textit{c} are the lattice constants. (e) Schematic of the LL transitions of 2D MDF in BaFe$_{2}$As$_{2}$. The green lines show the magnetic field dependence of the assumed \emph{E}$_{F}$. (f) Left panel: LLs of 3D MDF. Right panel: DOS of 3D MDF. (g) Calculated relative real part of the optical conductivity of 3D MDF based on Eq. (15) of Ref. \cite{Carbotte}. (h) Re[$\sigma$$_{xx}$(\emph{B})] $-$ Re[$\sigma$(0 T)] of BaFe$_{2}$As$_{2}$ at \textit{B} = 7 T. (i) Left panel: LLs of 2D MDF. Right panel: Lorentz-shaped DOS of 2D MDF. (j) Calculated Re[$\sigma$$_{xx}$(\emph{B})] $-$ Re[$\sigma$(0 T)] of 2D MDF based on Eq. (23) of Ref \cite{Carbotte1}. When calculating the spectra in (g) and (j), we replaced the Kronecker delta functions in Eq. (15) of Ref. \cite{Carbotte} and Eq. (23) of Ref \cite{Carbotte1} by the Lorentzian functions with non-zero scattering rates. 
	}
	\label{Fig:linear}
\end{figure}

To identify the nature of the LL transitions in BaFe$_{2}$As$_{2}$, we need to extract the absorption energies of the LL transitions from the measured reflectance spectra. Using the magneto-optical model to fit the absolute reflectance \emph{R}(\emph{B}) at a fixed magnetic field\cite{Palik} (see the grey curves in Fig. 1(b) and the upper panel of Fig. 1(c)), we obtained the dielectric function tensor $\varepsilon_{\pm}(B, \omega)$ for the right and left circularly polarized light and then got the real part of the diagonal optical conductivity Re[$\sigma$$_{xx}$(\emph{B}, $\omega$)] \cite{Suppl}. Figure 1(c) shows the obtained Re[$\sigma$$_{xx}$(\emph{B} = 17.5 T, $\omega$)] of BaFe$_{2}$As$_{2}$. Compared with the Re[$\sigma$(\emph{B} = 0 T, $\omega$)] of BaFe$_{2}$As$_{2}$ in the AFM state, Re[$\sigma$$_{xx}$(\emph{B} = 17.5 T, $\omega$)] has two additional peak-like features: T$_{1}$ and T$_{2}$, which are better resolved in Re[$\sigma$$_{xx}$(\emph{B} = 17.5 T, $\omega$)] $-$ Re[$\sigma$(\emph{B} = 0 T, $\omega$)] (see Fig. 1(d)). These two peak-like features, which move towards higher energies with increasing the magnetic field, correspond to the absorption modes of the LL transitions. Following the definition in the magneto-optical study of BiTeI\cite{Tokura}, we assign the peak positions in Re[$\sigma$$_{xx}$(\emph{B} = 17.5 T, $\omega$)] $-$ Re[$\sigma$(\emph{B} = 0 T, $\omega$)] to the LL transition energies.

We plotted the T$_{1}$ and T$_{2}$ energies as a function of $\sqrt{B}$ in Fig. 2(a). Both T$_{1}$ and T$_{2}$ energies exhibit a linear dependence on $\sqrt{B}$ and the zero intercepts at \emph{B} = 0 T under linear extrapolations. Moreover, the T$_{2}$ and T$_{1}$ energies scale as 2.4 $:$ 1 (see the inset of Fig. 2(a)). The $\sqrt{B}$-dependence of the LL transition energies and the zero-energy intercepts at \emph{B} = 0 T are two signatures of MDF\cite{Tokura,Potemski,ZJiang,nlwang3}. To identify the dimensionality of MDF in BaFe$_{2}$As$_{2}$, we performed DFT+DMFT calculations \cite{Suppl,Yin1,Yin2}. In Fig. 2(b)-(d), our calculations show that in the AFM state, the band dispersions near $E_F$ are linear along $k_x$ and $k_y$ directions, but extremely weak along $k_z$ direction, which theoretically suggests 2D MDF in BaFe$_{2}$As$_{2}$. According to the effective Hamiltonian describing 2D MDF in PCIS\cite{Tohyama}, we can obtain (i) the LL energy spectrum without considering Zeeman effects \cite{Suppl}:
{\setlength\abovedisplayskip{5pt plus 5pt minus 5pt}
\setlength\belowdisplayskip{5pt plus 5pt minus 5pt}
\begin{eqnarray}
E_n^{2D} = \text{sgn}(n) \sqrt{2e\hbar v_{F}^2|n|B}
\end{eqnarray}
}where the integer \emph{n} is LL index, sgn(\emph{n}) is the sign function, $\upsilon_F$ is the effective Fermi velocity of LLs, \emph{e} is the elementary charge and $\hbar$ is Planck's constant divided by 2$\pi$, and (ii) the selection rule for the allowed optical transitions from LL$_n$ to LL$_{n'}$ \cite{Suppl}:
{\setlength\abovedisplayskip{5pt plus 5pt minus 5pt}
\setlength\belowdisplayskip{5pt plus 5pt minus 5pt}
\begin{eqnarray}
\Delta n = |n|-|n'|=\pm1.
\end{eqnarray} According to Eq. (1), the energies of the allowed optical transitions from LL$_n$ to LL$_{n'}$ are given by:
{\setlength\abovedisplayskip{5pt plus 5pt minus 5pt}
\setlength\belowdisplayskip{5pt plus 5pt minus 5pt}
\begin{eqnarray}
E_{n\rightarrow{n'}}^{2D}= \text{sgn}(n') \sqrt{2e\hbar v_{F}^2|n'|B}-\text{sgn}(n)\sqrt{2e\hbar v_{F}^2|n|B}
\end{eqnarray}Indeed, the LL transition energies of 2D MDF exhibit a linear dependence on $\sqrt{B}$ and converge to zero at \textit{B} = 0 T. More importantly, equation (2) and (3) indicate that two groups of optical transitions between the LLs of 2D MDF have the energy ratios which are quite close to that between T$_{1}$ and T$_{2}$: (i) ($\sqrt{2}$$-$1) $:$ 1 $\approx$ 1 $:$ 2.414 for the LL transitions,
LL$_{-2}$ $\rightarrow$ LL$_{-1}$ and LL$_{-1}$ $\rightarrow$ LL$_{0}$ (or LL$_{+1}$ $\rightarrow$ LL$_{+2}$ and LL$_{0}$ $\rightarrow$ LL$_{+1}$),
and (ii) 1 $:$ (1+$\sqrt{2}$) $\approx$ 1 $:$ 2.414 for the LL transitions, LL$_{-1}$ $\rightarrow$ LL$_{0}$ and LL$_{-1}$ $\rightarrow$ LL$_{+2}$ (or LL$_{0}$ $\rightarrow$ LL$_{+1}$ and LL$_{-2}$ $\rightarrow$ LL$_{+1}$). For case (i) (see the two grey arrows in Fig. 2(e)), we can expect: (1) $\upsilon_F$ $\approx$ 2.85 $\times$ $10^{5}$ m/s, calculated based on Eq. (3), (2) $E_F$ (see the green line in Fig. 2(e)) should be pinned on LL$_{+1}$ (or LL$_{-1}$), even at \emph{B} = 17.5 T. (If $E_F$ was pinned on the LLs higher than LL$_{+1}$ (or LL$_{-1}$), e.g. LL$_{+2}$ (or LL$_{-2}$), LL$_{+1}$ (or LL$_{-1}$) should be fully occupied (or depleted). According to Pauli exclusion principle, LL$_{0}$ $\rightarrow$ LL$_{+1}$ (or LL$_{-1}$ $\rightarrow$ LL$_{0}$) would be blocked. If $E_F$ was pinned on LL$_{0}$, LL$_{-1}$ (or LL$_{+1}$) should be fully occupied (or depleted). Then LL$_{-2}$ $\rightarrow$ LL$_{-1}$ (or LL$_{+1}$ $\rightarrow$ LL$_{+2}$) would be blocked. Thus, if $E_F$ was not pinned on LL$_{+1}$ or LL$_{-1}$, one type of the LL transitions in case (i) would not occur. See Fig. S6 of the Supplemental Material \cite{Suppl}) and (3) the spectral weight of the assumed LL transition T$_{1}$ should decrease as the magnetic field increases, since LL$_{+1}$ (or LL$_{-1}$) is depopulated by the magnetic field\cite{Potemski}. In contrast, T$_{1}$ in Fig. 1(b) and (d) becomes more dominant with the enhancement of the magnetic field. Besides, based on the above $\upsilon_F$ and equation (1), the corresponding $E_F$ in BaFe$_{2}$As$_{2}$ is at least 179 meV, which is much higher than the measured Fermi energy of $\sim$ 1 meV by ARPES\cite{Richard}. Thus, T$_{1}$ and T$_{2}$ should not come from the LL transitions in case (i). Case (ii) (see the red and blue arrows in Fig. 2(e)) means that the system has been in the quantum limit at \textit{B} $\leq$ 5 T , which is consistent with the low $E_F$ measured by ARPES\cite{Richard}. By fitting the $\sqrt{B}$-dependence of the energies of T$_{1}$ and T$_{2}$ in case (ii) based on Eq. (3), we deduced $\upsilon_F$ $\approx$ 1.18 $\times$ $10^{5}$ m/s, which is comparable to the values reported by ARPES and transport measurements\cite{Richard,Tanigaki1}. Therefore, T$_{1}$ and T$_{2}$ arise from the LL transitions, LL$_{-1}$ $\rightarrow$ LL$_{0}$ and LL$_{-1}$ $\rightarrow$ LL$_{+2}$ (or LL$_{0}$ $\rightarrow$ LL$_{+1}$ and LL$_{-2}$ $\rightarrow$ LL$_{+1}$), respectively. 

When $E_F$ $\sim$ 0, 3D Dirac semimetals also have a $\sqrt{B}$ dependence of the LL transition energies at low magnetic fields \cite{Orlita1}. Further evidence is required to determine whether MDF in BaFe$_{2}$As$_{2}$ are 2D or 3D. The left panel of Fig. 2(f) shows that at magnetic fields, the band dispersions of 3D MDF are transformed into a series of one-dimensional LLs (or Landau bands), which disperse with the momentum component along the field direction. The LL spectrum of 3D MDF has the form: 
\begin{eqnarray}
E_n^{3D} = \pm\delta_{n,0}\upsilon_{z}\hbar k_{z} + \text{sgn}(n) \sqrt{2e\hbar v_{F}^2|n|B+{(\upsilon_{z}\hbar k_{z})}^2}
\end{eqnarray}
where $\delta_{n,0}$ is the Kronecker delta function, $k_{z}$ is the momentum along $\textit{z}$-axis, $\upsilon_z$ is the Fermi velocity along $k_{z}$ direction. In the right panel of Fig. 2(f), the zeroth LLs of 3D MDF disperse linearly with $k_{z}$ and thus do not have a singularity of the density of states (DOS), while the other LLs have the singularities of DOS at $k_{z}$ = 0 \cite{Orlita1}. Since optical absorptions are determined by the joint DOS of the initial and final state, figure 2(g), which was obtained by the theoretical calculations based on the model of the optical conductivity of 3D linear band dispersions (see Eq. (15) of Ref. \cite{Carbotte}), shows that the peak feature (i.e. T$_{1}$) of the zeroth-LL-related transition LL$_{-1}$ $\rightarrow$ LL$_{\pm0}$ (or LL$_{\pm0}$ $\rightarrow$ LL$_{+1}$) in Re[$\sigma$$_{xx}$(\emph{B})] $-$ Re[$\sigma$(0 T)] is \textit{weaker} than that (i.e. T$_{2}$) of the LL transition LL$_{-1}$ $\rightarrow$ LL$_{+2}$ (or LL$_{-2}$ $\rightarrow$ LL$_{+1}$). In contrast, the Re[$\sigma$$_{xx}$(\emph{B} = 7 T)] $-$ Re[$\sigma$(0 T)] of BaFe$_{2}$As$_{2}$ in Fig. 2(h) show that its peak feature (i.e. T$_{1}$) of the zeroth-LL-related transition LL$_{-1}$ $\rightarrow$ LL$_{0}$ (or LL$_{0}$ $\rightarrow$ LL$_{+1}$) is much stronger than its T$_{2}$ feature, indicating that MDF in BaFe$_{2}$As$_{2}$ should \textit{not} be 3D. On the contrary to the zeroth LLs of 3D MDF, the LL$_{0}$ of 2D MDF has a singularity of DOS at zero energy (see Fig. 2(i)). Thus, in Fig. 2(j), the theoretical calculations based on Eq. (23) of Ref. \cite{Carbotte1} reveal that the peak feature (i.e. T$_{1}$) of the LL$_{0}$-related transition of 2D MDF is stronger than the T$_{2}$ feature of 2D MDF, which is consistent with our magneto-infrared results of BaFe$_{2}$As$_{2}$. The relative intensities of T$_{1}$ and T$_{2}$, combined with the calculated band dispersions along $k_{x}$, $k_{y}$ and $k_{z}$ directions, indicate 2D MDF in BaFe$_{2}$As$_{2}$.

\begin{figure}
\includegraphics[width=8.5cm]{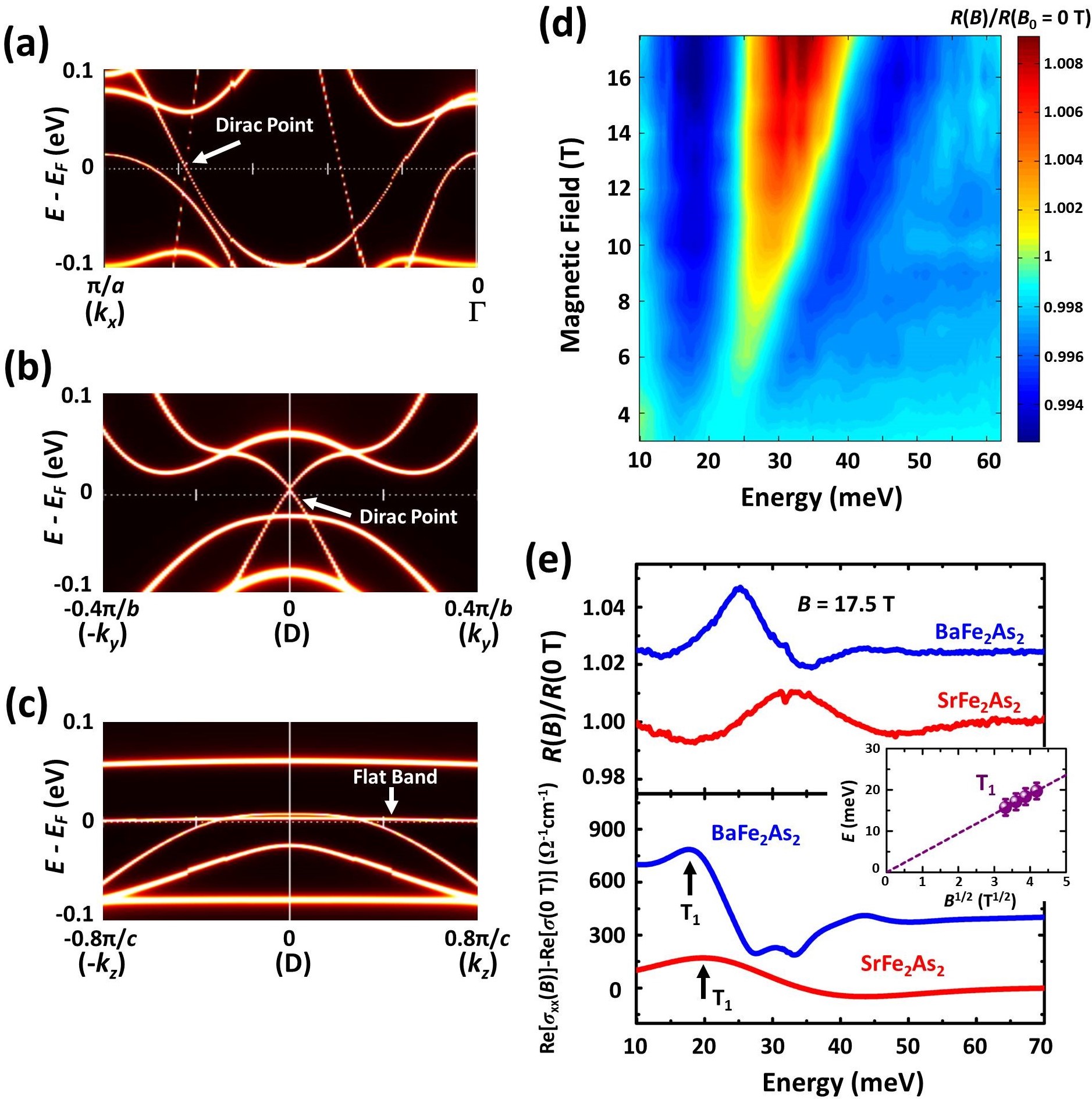}\\
\centering
\caption{
Magneto-optical response and band dispersions of SrFe$_{2}$As$_{2}$ in the AFM state. (a-c) Energy-momentum dispersions of MDF in SrFe$_{2}$As$_{2}$ along $k_x$, $k_y$ and $k_z$. The flat band in (c) shows the dispersion of MDF along $k_z$. (d) False-color map of the \emph{R}(\emph{B})/\emph{R}($B_0$ = 0 T) spectra of SrFe$_{2}$As$_{2}$ as a function of \emph{B} and \emph{E}. (e) Relative reflectance (upper panel) and relative optical conductivity (lower panel) spectra of BaFe$_{2}$As$_{2}$ and SrFe$_{2}$As$_{2}$ at \emph{B} = 17.5 T. The inset shows the $\sqrt{B}$-dependence of the observed T$_{1}$ transition energy. The spectra of BaFe$_{2}$As$_{2}$ in (e) are displaced from those of SrFe$_{2}$As$_{2}$ by 0.023 and 412 ($\Omega$$^{-1}$ cm$^{-1}$) for clarity.
}
\label{Fig:linear}
\end{figure}

To test the robustness of the Dirac points in PCIS, we completely replaced the barium atoms in BaFe$_{2}$As$_{2}$ by strontium atoms. The element replacement causes a crystal distortion, but (1) the AFM order still exists, and (2) inversion symmetry and the combination of time-reversal and spin-reversal symmetry are not broken\cite{Johnston}. In Fig. 3(a)-(c), our DFT+DMFT calculations show the linear band dispersions along $k_x$ and $k_y$ directions and the quite weak dispersions along $k_z$ direction, indicating 2D MDF in SrFe$_{2}$As$_{2}$. We further performed magneto-reflectance measurements of SrFe$_{2}$As$_{2}$ single crystals at \emph{T} $\sim$ 4.5 K in Faraday geometry (see Fig. 3(d)). In Fig. 3(e), the peak features in the relative reflectance and relative optical conductivity spectra of SrFe$_{2}$As$_{2}$ at \emph{B} = 17.5 T, which arise from the optical transition LL$_{-1}$ $\rightarrow$ LL$_{0}$ (or LL$_{0}$ $\rightarrow$ LL$_{+1}$), appear at higher energies than those of BaFe$_{2}$As$_{2}$, implying a larger effective Fermi velocity of the LLs in SrFe$_{2}$As$_{2}$ (see the spectra at some other magnetic fields in Fig. S7 of the Supplemental Material\cite{Suppl}). For SrFe$_{2}$As$_{2}$, in the inset of Fig. 3(e), the energy positions of the peaks in its relative optical conductivity spectra show a linear dependence on $\sqrt{B}$ and converge to zero at zero-field, consistent with the DFT+DMFT calculation results about 2D MDF. Fitting the $\sqrt{B}$-dependence of the peak features of SrFe$_{2}$As$_{2}$ yields $\upsilon_F$ $\approx$ 1.30 $\times$ $10^{5}$ m/s, which is indeed larger than that of BaFe$_{2}$As$_{2}$. Our infrared study, combined with the DFT+DMFT calculations, manifests that the substitution of the barium atoms in BaFe$_{2}$As$_{2}$ by strontium atoms not only maintains 2D MDF, but also enhances their effective Fermi velocity, providing evidence for the topologically protected Dirac points in PCIS.

In summary, our observation of the $\sqrt{B}$-dependence of the LL transition energies, the zero-energy intercept at \emph{B} = 0 T under linear extrapolations of the transition energies, the energy ratio ($\sim$ 2.4) and the dominant absorption features of the zeroth-LL-related transitions, together with the linear band dispersions in 2D momentum space obtained by DFT+DMFT calculations, not only demonstrates the existence of 2D MDF, but also provides support for the topological protection of the Dirac points in the AFM states of two iron-arsenide parent compounds---BaFe$_{2}$As$_{2}$ and SrFe$_{2}$As$_{2}$. 

\setlength{\parskip}{2.3em} 

\textit{Acknowledgements.} We thank Nanlin Wang, Marek Potemski, Zhiqiang Li, Milan Orlita, Ying Ran, Oskar Vafek and Pierre Richard for very helpful discussions. The authors acknowledge support from the National Key Research and Development Program of China (Project No. 2017YFA0304700, 2016YFA0300600 and 2016YFA0302300), the Pioneer Hundred Talents Program of Chinese Academy of Sciences, the National Natural Science Foundation of China (Grant No. 11674030), the start-up funding 10100-310432102 of Beijing Normal University, the European Research Council (ERC ARG MOMB Grant No. 320590) and the U.S. Department of Energy, Office of Basic Energy Sciences, under Contract No. DE-AC02-98CH10886. The single crystal and materials work at Rice is supported by the U.S. DOE, BES under contract No. \textrm{DE-SC}0012311 (P.D.). A portion of this work was performed in NHMFL which is supported by National Science Foundation Cooperative Agreement No. DMR-1157490 and the State of Florida. The DFT+DMFT calculations used high performance cluster at the National Supercomputer Center in Guangzhou.

\setlength{\parskip}{1em} 
\textit{Author Contributions.} Z.-G.C. conceived this project, carried out the optical experiments, analyzed the data and wrote the paper; L.W. calculated the LL spectrum and the selection rules; Z.Y., K.H. and G.K. did the DFT+DMFT calculations; Y.S., X.L., H.L., C.Z. and P.D. grew the single crystals.


\end{document}